\documentclass[12pt]{article}
\usepackage{amsmath,amssymb,epsfig}
\usepackage{graphicx,floatflt,subfigure}
\usepackage{epstopdf}

\usepackage{amsmath,caption,booktabs}
\usepackage{color}
\input{colordvi.tex}
\usepackage[table]{xcolor}

\makeatletter

\renewcommand\section{\@startsection {section}{1}{\z@}%
                                   {-3.5ex \@plus -1ex \@minus -.2ex}%
                                   {2.3ex \@plus.2ex}%
                                   {\normalfont\large\bfseries}}

\renewcommand\subsection{\@startsection{subsection}{2}{\z@}%
                                     {-3.25ex\@plus -1ex \@minus -.2ex}%
                                     {1.5ex \@plus .2ex}%
                                     {\normalfont\normalsize\bfseries}}

\makeatother

\begin{document}

\baselineskip=18pt  
\numberwithin{equation}{section}  
\allowdisplaybreaks  



%
%


\thispagestyle{empty}

\vspace*{-2cm}
\begin{flushright}
\end{flushright}

\begin{flushright}
\end{flushright}

\begin{center}

\vspace{2.8cm}

{\bf\Large Instability of Higgs Vacuum via String Cloud}
\vspace*{0.2cm}

\vspace{1.3cm}

{\bf
Issei Koga$^{1}$, Sachiko Kuroyanagi$^{2,3}$ and Yutaka Ookouchi$^{4,1}$} \\
\vspace*{0.5cm}

${ }^{1}${\it Department of Physics, Kyushu University, Fukuoka 819-0395, Japan  }\\

${ }^{2}${\it Department of Physics, Nagoya University, Nagoya 464-8602, Japan }\\

${ }^{3}${\it Instituto de F\'isica Te\'orica, Universidad Auton\'oma de Madrid, 28049 Madrid, Spain  }\\

${ }^{4}${\it Faculty of Arts and Science, Kyushu University, Fukuoka 819-0395, Japan  }\\

\end{center}

\vspace{2cm} \centerline{\bf Abstract} \vspace*{0.5cm}

We study the instability of the Higgs vacuum caused by a cloud of strings. By catalysis, the decay rate of the vacuum is highly enhanced and, when the energy density of the cloud is larger than the critical value, a semi-classical vacuum decay occurs. We also discuss the relation between the string cloud and observational constraints on the cosmic strings from the viewpoint of the catalysis, which are converted into bounds on the parameters of the Higgs potential.

\newpage
\setcounter{page}{1} 



\section{Introduction}

The discovery of the Higgs particle and the precise measurements of the top quark mass seems to reveal that our vacuum, in which the electroweak symmetry breaks down (we call the Higgs vacuum for short), is metastable \cite{InstabilityHiggs}. This fact has been boosting studies on the Higgs vacuum from various point of views \cite{Moroi}. These decay processes are known as the homogeneous vacuum decay. On the other hand, the inhomogeneous vacuum decay, initiated by \cite{Pole}, can occur in nature. The idea was later applied to phenomenological model building \cite{Pheno} and the vacuum decay in string theories \cite{Ookouchi} and gravity theories \cite{Hiscock,Gregory1401,Gregory1503,Gregory,Oshita}. Among them, the black hole catalysis discussed in \cite{Hiscock,Gregory1401} is interesting because it is generally applicable to various settings; The catalysis seeded by a topological soliton \cite{Pole,Pheno} highly depends on the structure of the potential. Typically, the soliton is stabilized by the topological charge related to the symmetry breaking. For the catalysis to work in this case, the true vacuum has to be connected to the symmetry restoring point of the potential as emphasized in \cite{Pheno}.  

The catalytic effects caused by the string cloud \cite{Cloud} was recently discussed in the context of the creation of the bubble Universe in five dimensions proposed in \cite{Danielsson}. The catalysis provided a kind of the selection rule to the cosmological constant \cite{KogaOokouchi}. 
In this paper, with the aim of getting interesting phenomenology, we apply the method to the decay of the Higgs particle in the standard model, basically along the lines of \cite{Gregory1503}. We then study the relation between the cloud of string and cosmological observations by showing how to connect the Higgs potential with the cosmic string tension. If the cloud of strings exists in nature, it can leave signatures in the cosmic microwave background (CMB) and gravitational wave probes by pulsar timing arrays and laser interferometers.
Assuming that the string cloud network behaves as the standard scaling cosmic string network, we apply cosmological bounds on the tension to obtain constraints on parameters of the Higgs potential. We also consider how future gravitational wave observations can help to test the scenario of Higgs vacuum decay through the catalytic effect of the string cloud. On the contrary, if the Higgs potential parameters are determined by future collider experiments, one can infer the string tension, which can be used mutually with cosmological observations. 

The organization of this paper is as follows. In section 2, we briefly review the method discussed in \cite{Coleman,CDL,Hiscock,Gregory1401} and apply to the decay of de Sitter to anti-de Sitter (AdS) vacua catalyzed by a cloud of strings \cite{Cloud} in the aim of application to the Higgs vacuum decay. We show that there exists the critical value above which semi-classical decay occurs rather than quantum tunneling. In section 3, we discuss the instability of the Higgs vacuum. In section 4, we investigate a connection between a string cloud and cosmological constraints on cosmic strings. From this, we study constraints for the parameters of Higgs potential. The section 5 is devoted to conclusions.

\section{The catalytic decay of de Sitter vacua }

In this section, first we review the general study on the catalytic decay of vacua discussed in \cite{CDL,Hiscock,Gregory}, and then we apply it to the cloud of strings. We compute the bounce action for the decay of a de Sitter vacuum to the Minkowski vacuum and compare with that of Coleman-de Luccia \cite{CDL}.

Consider a cloud of strings in a four-dimensional spacetime with a cosmological constant $\Lambda$. It is constructed of the relativistic strings and by smearing the energy density, one can get the cloud of strings.  We consider a node emanating several legs, giving rise to spherically symmetric cloud of strings. The solution for the Einstein equation of the spacetime is given by \cite{Cloud}
\begin{equation}
ds^2= -f(r) dt^2 +{dr^2 \over f(r)}+r^2\left(d \theta^2+\sin^2 \theta d \phi^2 \right)~, \label{cloudsolution}
\end{equation}
where 
\begin{equation}
f(r)=1-{ \Lambda \over 3}r^2 -{a}~.
\end{equation}
Here, $a$ is essentially the tension of the cloud of strings. 

Now, we study a junction of two solutions with different cosmological constants and parameters. By using the subscript $+$ $(-)$ for quantities outside (inside) the wall, the junction conditions, known as the Israel's conditions \cite{Israel}, are described by  
\begin{eqnarray}
 \frac{1}{R}(f_+\dot \tau_+-f_-\dot \tau_-)&=&-{4\pi G}\sigma~,\label{Israel1}\\
  f_\pm\dot \tau_\pm^2+\frac{\dot R^2}{f_\pm}&=&1~,
\end{eqnarray}
where we introduced the Euclidean time $\tau$ by the Wick rotation, $t=-i\tau $. $R$ corresponds to the wall trajectory and the metric on the wall has the following form, 
\begin{equation}
ds^2= - d \lambda^2 +R^2(\lambda) \left(d \theta^2+\sin^2 \theta d \phi^2 \right)~. 
\end{equation}
From \eqref{Israel1}, the equation for the wall trajectory is given by 
\begin{equation}
\dot{R}^2=-\bar{\sigma}^2 R^2 +\bar{f}-{(\Delta f)^2 \over 16 \bar{\sigma}^2 R^2}~, \label{walltrajectory}
\end{equation}
where we defined
\begin{eqnarray}
  \bar f=\frac{1}{2}(f_++f_-)\ , \qquad \Delta f=f_+-f_- ~.
\end{eqnarray}
Adopting the same notation as \cite{Gregory1401,Gregory1503}, we introduce $\eta=\bar{\sigma}l$, $\bar{\sigma}={2\pi G \sigma }$ and
\begin{equation}
l^2={3 \over \Delta \Lambda} ,\qquad \gamma= {4\bar{\sigma}l^2 \over 1+4\bar{\sigma}^2 l^2},\qquad \alpha^2 =1+{\Lambda_- \gamma^2 \over 3}~. \label{parameters}
\end{equation}
Also, we define the dimensionless parameters by  
\begin{equation}
\tilde{R}={\alpha R\over \gamma}\  , \qquad \tilde{\lambda}={\alpha \lambda \over \gamma}~\, \qquad \tilde{\tau}={\alpha \tau \over \gamma}~. \label{normalization}
\end{equation}
From the explicit metric of the cloud \eqref{cloudsolution}, the equation of the wall reduces to 
\begin{equation}
\left({d \widetilde{R} \over d \tilde{\lambda} } \right)^2=1-\left( \widetilde{R} +  {k^{\prime }\over \widetilde{R}}\right)^2 - k ~, \label{Rdot2eq}
\end{equation}
where we defined
\begin{eqnarray}
k=   \left( a_-+{\Delta a (1-\alpha) \over 2\bar{\sigma}\gamma} \right)\ ,\qquad k^{\prime}={\Delta a \over  4\bar{\sigma} }\left( {\alpha \over \gamma} \right) ~,
\end{eqnarray}
and $\Delta a =a_+-a_-$.

Now we are ready to study the decay process of de Sitter space-time with $\Lambda_+>0$ to the anti de Sitter spacetime $\Lambda_-<0$. For convenience, we introduce 
\begin{equation}
l_{\pm} =\sqrt{ 3\over \pm \Lambda_\pm  } \ , \qquad \delta={l_-\over l_+}~.
\end{equation}
By following the method\footnote{First, we solve the equation of motion for $R$, then substitute it in the action. By subtracting the action corresponding to the initial state, we obtain the bounce action $B$.} used by Coleman and de Luccia, we compute the bounce action $B$ to estimate the decay rate. The method to compute the bounce action for inhomogeneous decay was recently developed by Gregory, Moss and Withers \cite{Gregory1401}. We will proceed the analysis basically along the lines of the paper. The action is divided into two parts, one is the contribution coming from the singularities of the bounce solution, while the other is from the regular part of the solution.

By doing the same way as \cite{Gregory1401} (see the appendix for an review), the first one is given by the area of the horizon ${\cal A}_i$ 
\begin{equation}
I_{\cal B}=-{1\over 4G}\sum_i {\cal A}_i~.
\end{equation}
On the other hand, the second one is 
\begin{equation}
I= -{1\over 2G}\left({\gamma \over \alpha} \right)^2 \int_{\tilde{\lambda}_{\rm min}}^{\tilde{\lambda}_{\rm max}} d \tilde{\lambda} \widetilde{R}^2\left[ \left({d f_+\over d \widetilde{R}}-{2f_+\over \widetilde{R}} \right){\dot{\widetilde{\tau}}}_+ - \left({d f_- \over d \widetilde{R}}-{2f_-\over \widetilde{R}} \right) \dot{\widetilde{\tau}}_- \right] ~,
\end{equation}
where $\lambda_{\rm min}$ and $\lambda_{\rm max}$ are obtained by the condition $\dot{R}=0$ as follows,
\begin{eqnarray}
\sqrt{2} \widetilde{R}_{\rm max}&=&\sqrt{2}\widetilde{R}(\tilde{\lambda}_{\rm max})=\left(  {1-k -2k^{\prime}+ \sqrt{(k-1)(k-1+4k^{\prime})}}   \right)^{1\over 2}~, \\
\sqrt{2} \widetilde{R}_{\rm min}&=&\sqrt{2}\widetilde{R}(\tilde{\lambda}_{\rm min})=\left(  {1-k -2k^{\prime}- \sqrt{(k-1)(k-1+4k^{\prime})}}   \right)^{1\over 2} ~.
\end{eqnarray}

\begin{figure}[t]
    \centering
    \includegraphics[width=7cm]{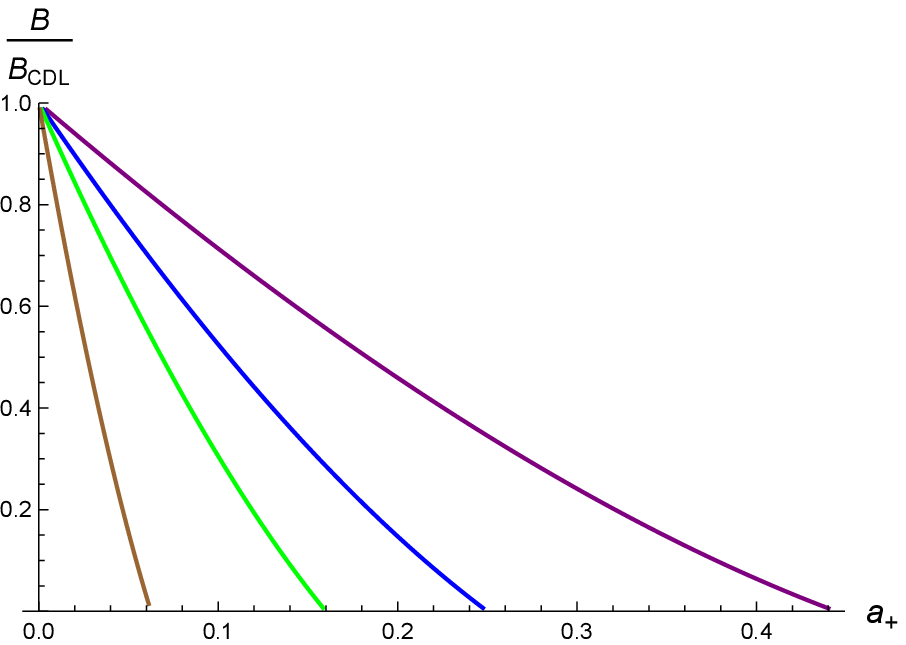}
    \includegraphics[width=7cm]{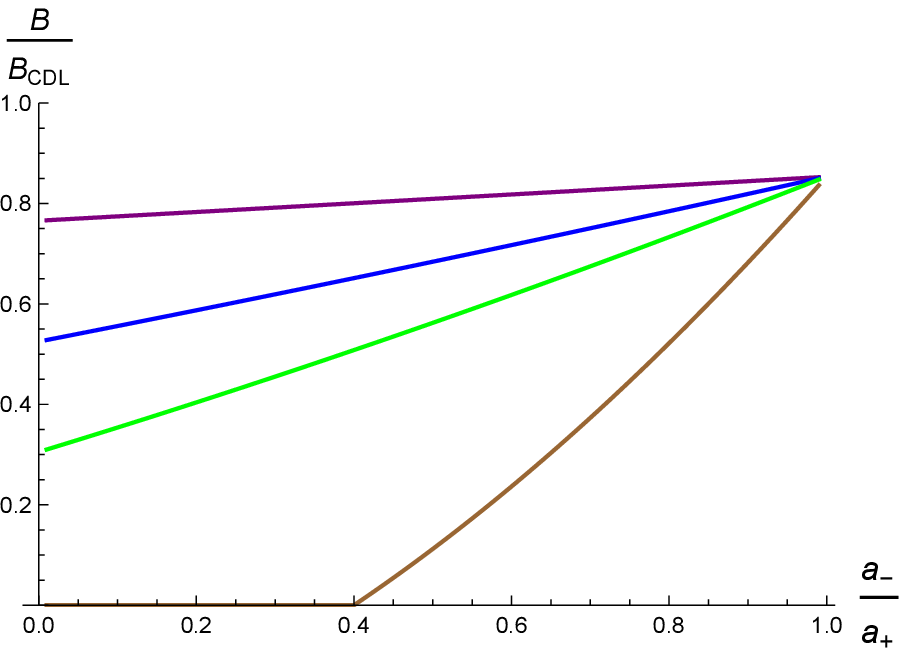} 
    \caption{\sl Left panel: The catalytic decay of de Sitter vacuum to anti-de Sitter vacuum. The bounce action for $\delta={l_-/ l_+}={1/10^4}$, $a_-=0$ and various choices of $\eta$.  From the bottom to the top, we choose $\eta=1/8$, $1/5$, $1/4$ and $1/3$. Right panel: The bounce action for the fixed $a_+=1/10$ and $\delta=1/10$.  We choose $\eta=1/8$, $1/5$, $1/4$ and $1/3$ from the bottom.}
    \label{Fig1}
\end{figure}

\begin{figure}[htbp]
    \centering
    \includegraphics[width=7cm]{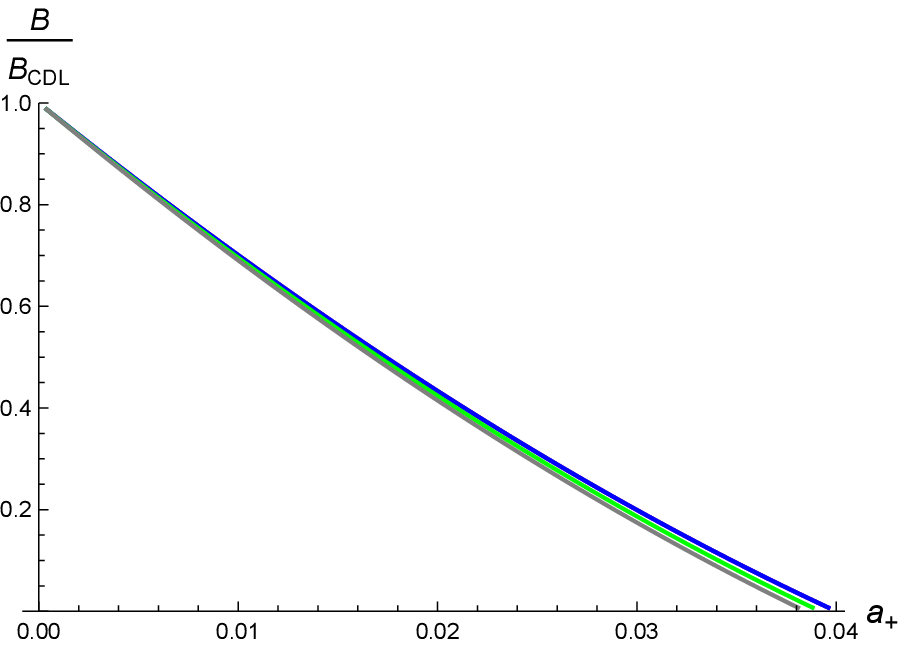}
    \includegraphics[width=7cm]{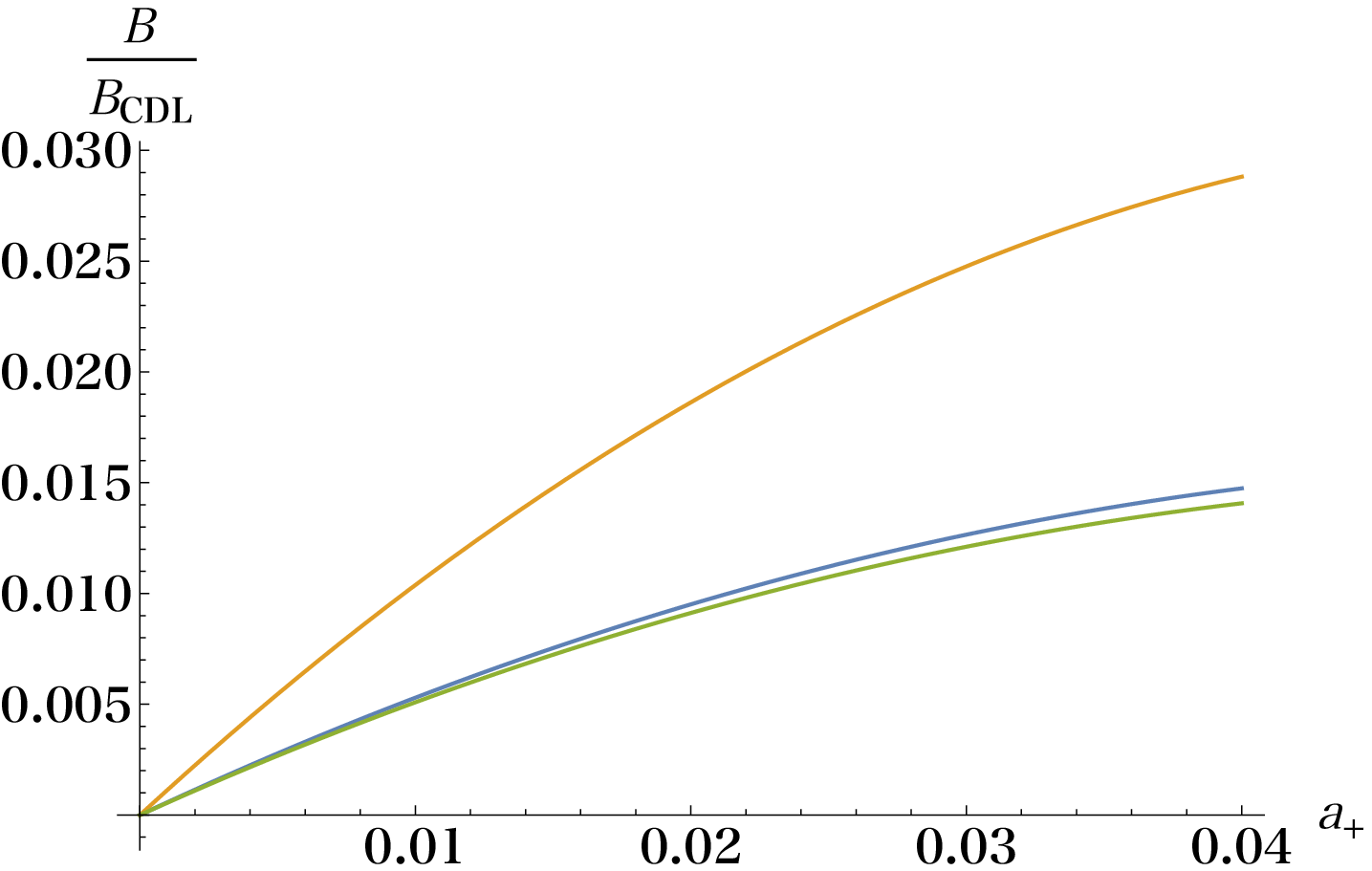} 
    \caption{\sl Left panel: The bounce action for $\eta=1/10$, $a_-=0$ and several choices of $\delta$. The blue curve corresponds to that for $1/10^4$, green for $1$, and gray for $10^4$. The difference of $\delta$ does not affect much. Right panel: We show differences between two curves. The top curve corresponds to the difference of the bounce actions for $\delta =1/10^4$ and $1$. The second and third curves correspond to those for $\delta=1/10^4$ and $10^4$ and $\delta =1$ and $10^4$. Clearly, compared to the values of bounce actions shown in the left panel, the differences are very small. }
    \label{Fig2}
\end{figure}

In computing the bounce action, we subtract the action corresponding to the initial state. The contribution from the cosmological horizon $r_c=\sqrt{3(1-a_+)/ \Lambda_+}$, which is the only singularity 
existing in the bounce solution in our setting, cancels out 
\begin{equation}
B=-{{\cal A}\over 4G}+I-\left( -{{\cal A}\over 4G}\right)=I~. \label{BounceAction}
\end{equation}
Hence the total is given by $I$. Below, we numerically compute this action with several choices of parameters. 
Remarkably, from Figure \ref{Fig1}, we find that there is a critical value above which the bounce action vanishes. This is in contrast to the catalysis by the black hole discussed in \cite{Gregory1401}. When the initial value of $a_+$ is larger than the critical value, the semi-classical vacuum decay happens where quantum tunneling is not required for the decay.   
From the right panel of Figure \ref{Fig1}, we see that the decay without remnants is dominant contribution for fixed $\eta$ and $a_+$. In Figure \ref{Fig2}, we show the bounce action for various choices of $\delta={l_-/l_+}$. From this we find that $\delta$ does not affect the bounce action much compared to those of $a_+$ and $\eta$.

To estimate the critical value, let us discuss the condition $R_{\rm max}=R_{\rm min}$, which yields 
$(k-1)(k-1+4k^{\prime})=0$. From this\footnote{We obtain two solutions, $a_+  = 8\eta^2/(1+4\eta^2) (1\pm \alpha)$. The smaller one gives stronger condition.}, we find that the critical value is given by
\begin{equation}
a_+^{(c)}={8\eta^2 \over 1+4\eta^2}(1+\alpha)^{-1}~. \label{criticalvalue}
\end{equation}
The decay with $a_+> a_+^{(c)}$ induces the semi-classical decay.

\section{The catalytic decay of Higgs Vacuum}

\begin{figure}[t]
    \centering
    \includegraphics[width=9.5cm]{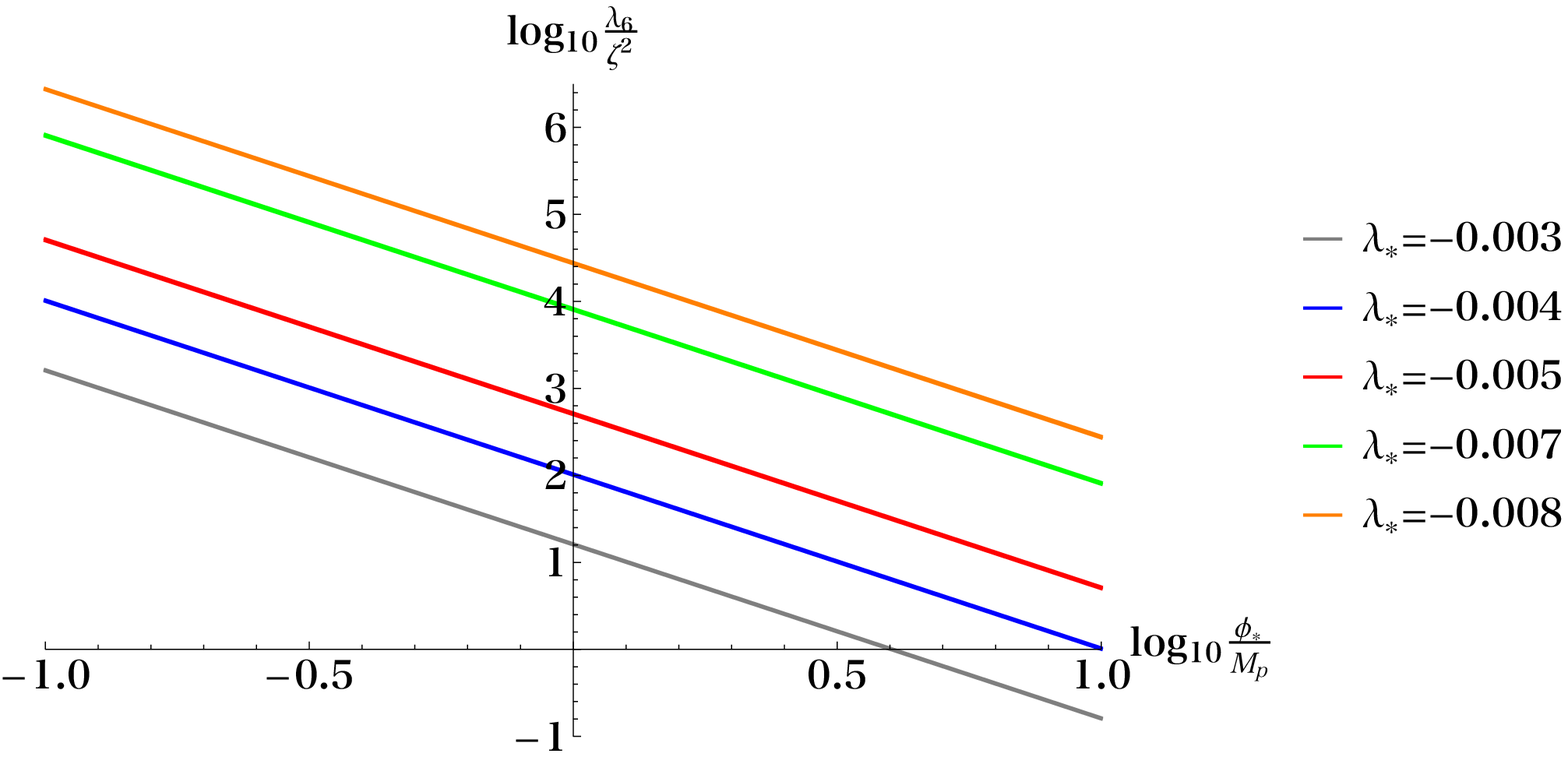}
     \includegraphics[width=7cm]{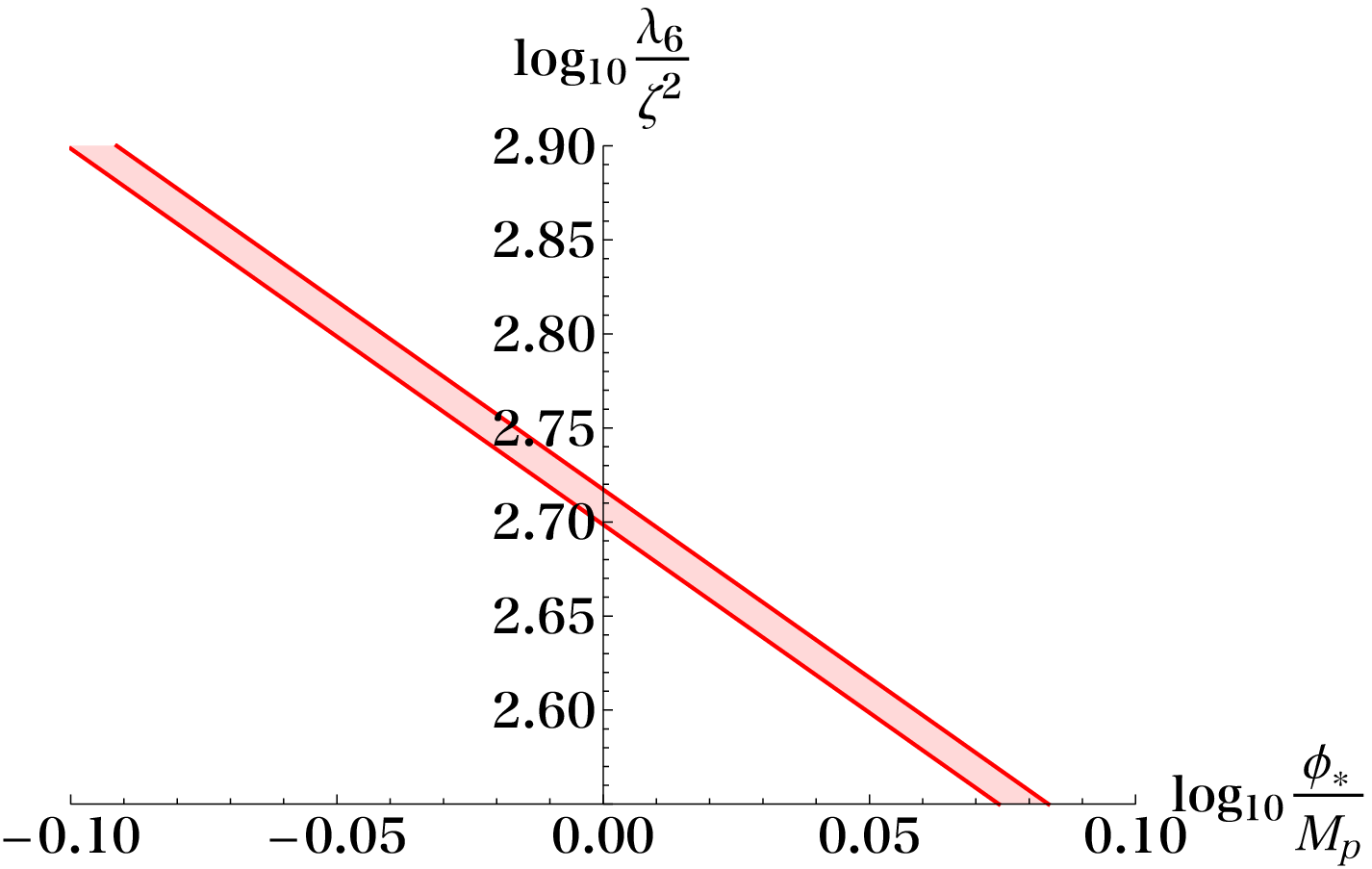} 
 \caption{\sl Left panel: Allowed parameter ranges for several choices of $\lambda_*$ in which the thin-wall approximation is valid. From the bottom, we took $\lambda_*=-0.003$, $-0.004$, $-0.005$, $-0.007$ and $-0.008$.  For each choice of $\lambda_*$, the allowed range is quite narrow, but by changing it, we can cover a large region of $\lambda_6$ - $\phi_*$ plane. We assumed $b=10^{-4}$ and ${(\delta \lambda)}_{\rm bsm}=\lambda_8=0$. Right panel: To emphasize the narrowness of the allowed region, we show the enlarged figure for $\lambda_*=-0.005$. }
    \label{FigB1}
\end{figure}

It is believed that the electroweak vacuum is metastable after recent measurements of the top quark mass and the discovery of the Higgs particle. In this section, we apply the method developed in the previous section to the decay of Higgs vacuum \cite{InstabilityHiggs}. In principle, we can use the precise two-loop order of the Higgs potential for this analysis, however it is quite involved. Thus, we here use a toy model of the potential that almost recover the Higgs potential. According to \cite{Kawai,Gregory1401}, we adopt the following potential,
\begin{equation}
V(\phi)={1\over 4}\lambda_{\rm eff}(\phi)\phi^4+{1\over 4}(\delta \lambda)_{\rm bsm}\phi^4+{\lambda_6 \over 6} {\phi^6\over M_{\rm new}^2}+\cdots ~,
\end{equation}
where 
\begin{equation}
\lambda_{\rm eff}(\phi )\simeq \lambda_* +b \left( \ln {\phi \over \phi_*}\right)^2~,
\end{equation}
with $b$ being a constant of order $10^{-4} \mathchar`- 10^{-5}$. The first term is the contribution within the standard model while the second and third contributions come from the beyond standard model. As the scale of the new physics we naively assume $M_{\rm new}=\zeta M_{\rm pl}$ with $\zeta\le 1$. The computations in the previous section rely on the thin-wall approximation. So we constrain the parameters in the potential to the range where thin-wall approximation is valid. Roughly speaking, when the peak of the potential is large enough compared to the depth of the true vacuum. As an illustration, in Figure \ref{FigB1} we show the parameter ranges for several choices of $\lambda_*$ where the thin-wall approximation is reliable. For each choice of $\lambda_*$, the allowed range is quite narrow, but adjusting it appropriately, we can cover a large region of $\lambda_6$ - $\phi_*$ plane.

\begin{figure}[htbp]
    \centering
    \includegraphics[width=9.5cm]{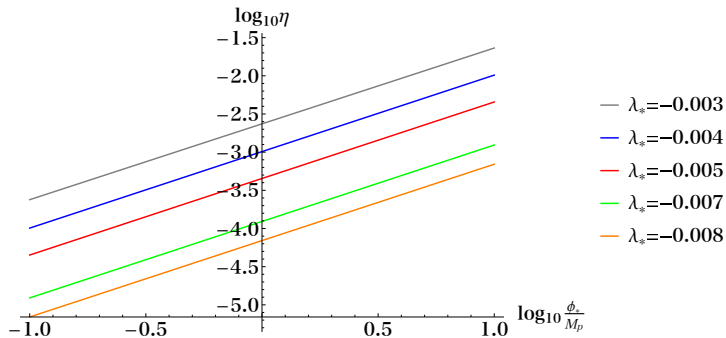}
    \caption{\sl $\eta$ for each choice of parameters of Higgs potential. We assumed $b=10^{-4}$ and ${(\delta \lambda)}_{\rm bsm}=\lambda_8=0$.  As for $\lambda_6/\zeta^2$, we used values in the allowed region of Figure \ref{FigB1}. Each color corresponds to that of Figure \ref{FigB1}.}
    \label{FigB2}
\end{figure}


In this case, the tension can be described by
\begin{equation}
\sigma=\int_{\phi_{\rm fv}}^{\phi_{\rm tv}}d\phi \sqrt{2\left[ V (\phi)- V_{\rm tv} \right] }~.
\end{equation}
The initial value of $\phi$ is the weak-scale which is much smaller than the Planck scale. So, we simply take $\phi_{\rm fv}=0$. Also, the cosmological constant at the present age is much smaller than the absolute value of that of the true vacuum, hence we set $\Lambda_+=0$ hereafter. In this case, since the parameter $\delta $ vanishes, namely $\alpha=(1-4\eta^2)/(1+4\eta^2)$, the critical value shown in \eqref{criticalvalue} becomes
\begin{equation}
a_+^{(c)}=4\eta^2={6\pi G \over -V_{\rm tv}}\left( \int^{\phi_{\rm tv}}_{\phi_{\rm fv}}d\phi \sqrt{2[V(\phi)-V_{\rm tv}]} \right)^2 ~,\label{CE}
\end{equation}
where we used $l_-=\sqrt{-3/8\pi GV_{\rm tv}}$. Hence, when the initial impurity is larger than $4\eta^2$, the semi-classical vacuum decay occurs. To estimate the order of the critical value, in Figure \ref{FigB2} we show $\eta=2\pi G \sigma l$ for each choice of parameters of Higgs potential by assuming $b=10^{-4}$ and ${(\delta \lambda)}_{\rm bsm}=\lambda_8=0$.

\section{Cosmological constraints on the cloud of strings}

In this section we study the relation between the cloud of strings and cosmological observations of cosmic strings. Since constraints on cosmic strings are typically provided in terms of the tension of strings, let us begin by relating the parameter $a_+$ introduced in the previous sections and the tension of strings. To do that, we compute the $(0,0)$-component of the Einstein equation by putting $\Lambda=0$ to extract a contribution of the string alone. By plugging the explicit metric \eqref{cloudsolution} into the Einstein equations, one obtains the total energy density
\begin{equation}
\rho_{\rm cl}(r)={c^2\over 8\pi G}{a\over r^2}~.\label{TensionCloud}
\end{equation}
Integrating it over a distance $L$ from the origin, we get the total energy $E$ stored in a sphere with radius $L$,
\begin{equation}
E=\int_0^{L} \rho_{\rm cl} \, 4\pi r^2={c^2 a \over 2 G}L~.
\end{equation}
The tension of the cloud of strings is obtained by dividing the energy $E$ by $L$,
\begin{equation}
\mu\equiv {E\over c^2 L}={a \over 2 G}~.\label{Gmu}
\end{equation}
Thus, the parameter $a_+$ of the string cloud solution is nothing but $2G\mu$. This is interesting because we can apply the cosmological constraints on cosmic strings to the catalytic vacuum decay, as we will see below. 

Consider the evolution of the cloud of strings without specifying the generation mechanism. In the standard scenario of cosmic strings (see \cite{DV04} for a review), from numerical simulations of the string evolution, it is believed that the string network reaches the scaling regime where the ratio $\gamma=\xi/t$ becomes asymptotically constant. $\xi$ is the typical scale of strings. When the initial density of strings is large, the interaction between strings enhances and the strings decay efficiently, which reduces the initial density. On the other hand, when the initial density is small, the interaction between them becomes rare. Also, by the expansion of the Universe, more strings come in from outside of the horizon, which eventually increases the string density. Hence, the details of the initial distribution do not matter, as long as there are some infinite strings. The energy density of cosmic strings is given by 
\begin{equation}
\rho_{\rm st}={\mu \over \xi^2}~.
\end{equation}
In the scaling regime, it goes as $\rho_{\rm st}\sim 1/t^2$ just like the total energy density $\rho_{\rm tot}=3H^2/(8\pi G)$, so that their ratio is constant and does not dominate the energy density of the Universe,
\begin{equation}
{\rho_{\rm st}\over \rho_{\rm tot}}={8\pi G\mu \gamma^2 \over 3\nu^2}\ll 1~,
\end{equation}
where we have used that the Hubble expansion rate is given by $H=\nu t$ with $\nu=1/2$ or $2/3$ for the radiation and matter era, respectively. 

Although the string cloud is slightly different from the standard cosmic string since it has a junction point of strings, one can naively expect that the network of the string cloud also reaches the scaling regime: In \cite{Vilenkin1,Martins}, the network of cosmic necklaces or cosmic lattices, which has junction points on strings where monopoles are attached to two or more strings, were studied and it was concluded that the string network reaches the scaling regime. It may be also similar to the network evolution of the cosmic strings attached to primordial black holes, which was discussed in \cite{Vilenkin2}. Naively speaking, when the mass of the junction point $m$ and the distance between junctions $d$ satisfy $m/\mu d \ll 1$, the contribution from the junction points to the network evolution is negligible. Since we assume that the junction point of the string cloud is massless, the condition is satisfied. Also, in \cite{McGraw}, the cosmic string network with Y-junctions, which is a slightly different setup but shares some features with our model, was discussed and the conclusion was that the network reaches the scaling behaviors as well. Hence, we simply assume that the network of the string cloud behaves like standard cosmic strings, and apply the constraint for the standard comic strings to our cloud. Note that even in the case junctions interrupt the string network to form loops and prevent the scaling regime, one may expect that strings intersect more often after the string density increases while the Hubble horizon grows and it eventually enhances the loop production and leads to the scaling solution. In this case, the number density of strings in the Universe becomes larger and cosmological constraints on the string tension get stronger. In this sense, our assumption gives conservative constraints. 

From the recent CMB observations by Planck \cite{Ade:2015xua}, cosmic strings with high scale tensions were ruled out, allowing us to assume strings with $G\mu < {\cal O} (10^{-7})$. Pulsar timing arrays probe gravitational waves at the nano-Hertz frequencies $10^{-9 } \mathchar`-10^{-8}$~Hz and have placed the strongest constraint so far, $G\mu \le  {\cal O}(10^{-10})$  \cite{Arzoumanian}. Note that the constraint by pulsar timing arrays changes depending on the assumed loop size distribution, and the conservative limit is $G\mu < {\cal O} (10^{-7})$ \cite{Lentati}. Advanced-LIGO observes gravitational waves at high frequency $\sim 10^2$~Hz and has given bounds on the string tension using results from both burst search \cite{LIGO1} and stochastic background search \cite{LIGO2}. The upper limit on $G\mu$ again depends on the assumed model of loop size distribution, but the conservative limit is $G\mu < {\cal O} (10^{-6})$. Hence we naively assume that the allowed range of the string is as follow: 
\begin{equation}
G\mu < 10^{-7}~.\label{observation}
\end{equation}

Now we are ready to apply this constraint to our analysis of Higgs vacuum decay by the catalysis. Since $a_+$ corresponds to $2G\mu$, the constraint \eqref{observation} immediately translates into that of the seed for the catalysis. As an illustration, let us take sample values for the Higgs potential and compute the numerical values of the critical point from \eqref{CE}. Table \ref{table1} is a list of the critical values $a_+^{(c)}$ for some sample parameter choices of the Higgs potential, which shows that $a_+^{(c)}$ is of order ${\cal O}(10^{-7} \mathchar`- 10^{-6})$. Comparing the constraint \eqref{observation}, we find that in the parameter region where the thin-wall approximation is valid, large initial values of the seed, $a_+\ge a_+^{(c)}$, are almost excluded. Remarkably, this is consistent with the long life-time of our Universe. If the semi-classical decay had happened because of the large value of $a_+$, then the Universe could have ended in the early stage.

\begin{table}
\centering
$\begin{array}{ ccccc }
\toprule
 \lambda_* & \phi_*/ M_{\rm pl} & \lambda_6 & \eta=2\pi G \sigma l &a_+^{(c)} \\
\midrule
-0.005 & 1 & 5\times 10^2 & 9.8\times 10^{-4} & 3.8\times 10^{-6} \\ 
 -0.007    & 0.52 & 3\times 10^{4} & 1.6\times 10^{-4} & 1.0\times 10^{-7} \\ 
 -0.008 & 1.16 & 2\times 10^4 & 7.6\times 10^{-5}          & 1.1\times 10^{-7}  \\ 
\bottomrule
\end{array}$
\caption{\sl Sample parameter choices of the Higgs potential and the corresponding critical values.}
\label{table1}
\end{table}

It is fascinating that future gravitational-wave experiments will reach the detectable sensitivity of lower scale tension. For example, the ground-based detector network, consisting of Advanced-LIGO, Advanced-VIRGO, and KAGRA, will improve the sensitivity to gravitational waves at $\sim 10^2$~Hz and will get access to $G\mu \sim 10^{-11}$ \cite{Kuroyanagi1}. The Square Kilometer Array will enhance the sensitivity pulsar timing array and be able to reach $G\mu \sim 10^{-12}$ \cite{Kuroyanagi2}. Furthermore, space-borne gravitational wave detectors such as LISA and DECIGO will probe gravitational waves at $\sim 10^{-3}$~Hz and $\sim 10^{-1}$~Hz with unprecedented sensitivity, which will enable us to go down to $G\mu \sim 10^{-17}$ \cite{Auclair} and $G\mu \sim 10^{-21}$ \cite{Kuroyanagi3}, respectively. If these experiments could detect a signal of cosmic strings, we would extract information on the Higgs potential through the catalysis by attributing the signal to the string cloud. Since the inhomogeneous vacuum decay is still dominant in this range, the Higgs vacuum may decay faster than we expect.

\section{Conclusions}

In this paper, we discussed the catalytic decay of the Higgs vacuum seeded by the cloud of strings. As an illustration, we computed the bounce action for the decay for the fixed $\delta$ and $\eta$ and found that, for sufficiently large energy scale of the cloud, the vacuum decay does not require the quantum tunneling and the semi-classical decay occurs instead. Even though the vacuum itself is long-lived, the short-time decay is enforced by the string cloud.  We found that the critical value crucially depends on the parameters of Higgs potential. This is interesting because the existence of our Universe until the present age suggests that some choices of the Higgs parameters are not allowed if one assume the network of the string cloud exists. We also showed that the Higgs parameters can be related to the tension of cosmic strings and the current upper bound on the string tension is consistent with the fact that the semi-classical decay has not happened in our Universe. In future, if we can observe the signal of cosmic strings and naively identify it with the string cloud, we can obtain unique constraints for the Higgs potential. 
%

\section*{Acknowledgments}

IK and YO are grateful to the Yukawa Institute for Theoretical Physics at Kyoto University, where the first stage of this work was done during the YITP-W-19-05 on ``Progress in Particle Physics 2019'' and YITP-W-19-10 on ``Strings and Fields 2019''. SK is partially supported by JSPS KAKENHI Number JP17K14282 and Career Development Project for Researchers of Allied Universities. YO is supported by JSPS KAKENHI Grant Numbers JP17K05419 and JP18H01214 and Qdai-jump Research Program of Kyushu University (No.01300).

\appendix

\section{Contributions from singular parts of the bounce solution}

In this appendix, we briefly review how to treat singular parts of the bounce solution and compute the bounce action along the lines of \cite{Gregory1401}. Roughly, the bounce solution corresponds to a motion stating from an unstable point of the Euclideanized potential and bounce back by the potential barrier and come to the original position. Hence, it has the period of the motion which we denote $T$. On the other hand, the solution has a horizon where $f(r_h)=0$. Near the horizon, it is convenient to introduce the coordinate $\rho$ defined by  $d \rho = {d r / \sqrt{f}}
$, in which the horizon places at the origin. The metric \eqref{cloudsolution} can be written as follows;
\begin{equation}
d s^2 =f(r(\rho))d \tau^2 + d\rho^2 +r(\rho)^2 \left( d \theta^2 +\sin^2 \theta d \phi^2 \right)~.
\end{equation}
The period $T$ of the bounce solution is not $2\pi$, in general. Thus, we introduce new time variable $\chi$ having $2\pi $ periodicity. By representing the $\tau$ in terms of the new variable, the metric reduces to    
\begin{eqnarray}
d s^2 =F(\rho)^2 d \chi^2 + d\rho^2 +r(\rho)^2 \left( d \theta^2 +\sin^2 \theta d \phi^2 \right)~,\label{metric2}
\end{eqnarray}
where we defined 
\begin{equation}
F(\rho)^2 \equiv f\big(r(\rho)\big)\left({T \over 2\pi} \right)^2 ~.
\end{equation}
Consider the behavior of the function near the horizon $\rho\simeq 0$. Since $F$ is proportional to $f(r)$ with a positive power, it vanishes in the limit $\rho \to 0$. So, the leading contribution in $\rho$ is ${\cal O}(\rho)$. On the other hand, the function $r(\rho)$ goes to $r_h$ in the limit. From the definition of $\rho$, we find that the derivative of $r$ with respect to $\rho$ vanishes, 
\begin{equation}
{d r\over d \rho }=\sqrt{f}\to 0 \ , \quad (r\to r_h)~.
\end{equation}
In all, near the origin the functions behave as follows;
\begin{equation}
F\simeq \rho F^{\prime}(0) \ , \qquad r(\rho)\simeq r_h+ {1\over 2}r^{\prime \prime}(0)\rho^2~.
\end{equation}
Substituting these expressions for \eqref{metric2}, we obtain
\begin{equation}
d s^2=d \rho^2+\rho^2 d \left( F^{\prime}(0)\chi \right)^2+\cdots ~.
\end{equation}
If $F^{\prime}(0)\chi$ has $2\pi$ periodicity, there is no deficit angle. However, since $F^{\prime}(0)\neq 1$ in general, the deficit angle $\delta$ defined by the following expression exists,
\begin{equation}
\delta =1-{F(\rho)\over \rho}\Big|_{\rho \to 0}=1-F^{\prime}(0)~.
\end{equation}

Now, let us review the computation of the bounce action from the singularities. We denote the vicinity of the singularities ${\cal B}=\sum_i {\cal B}_i$ and decompose the action into two parts, $I= I_{ {\cal M}-{\cal B} } +I_{\cal B}$. Each term include a boundary, hence we add the Gibbons-Hawking terms,
\begin{eqnarray}
I_{ {\cal M}-{\cal B} }&=&-\frac{1}{16\pi G_D} \int_{{\cal M}-{\cal B}} R -\int_{{\cal M}-{\cal B}}{\cal L}_m+{1\over 8\pi G_D}  \int_{\partial ({\cal M}-{\cal B})}K ~, \\
I_{ {\cal B} }&=& -{1\over 16\pi G_D} \int_{\cal B} R -\int_{{\cal B}}{\cal L}_m+\frac{1}{8\pi G_D} \int_{\partial {\cal B}} K~. \label{Singular} 
\end{eqnarray}

To estimate the contributions from the singularities, we have to regularize them by replacing the vicinity ${\cal B}$ with an manifold without singularities. However we preserve the behavior near the cut the same. Suppose a singular part is cut at $\rho=\epsilon$ and manifold without deficit angle at the origin is glued. In this case, the metric near the singularity is slightly modified and the function $F$ should be different in the region $\rho <\epsilon$. So we introduce new function $\widetilde{F}(\rho)$. Since the singularity is removed, it should behave as $\widetilde{F}^{\prime}(\rho=0)=1$ at the origin. Also, at the cut $\rho=\epsilon$, it has to have the same behavior as before, the function satisfies
\begin{equation}
1-\delta ={\widetilde{F}(\epsilon)\over \epsilon}~.
\end{equation}
Hence, we obtain $\widetilde{F}^{\prime}(\epsilon)=1-\delta$. With this regularized metric, 
\begin{eqnarray}
d s^2 =\widetilde{F}(\rho)^2 d \chi^2 + d\rho^2 +r(\rho)^2 d \Omega ^2~,
\end{eqnarray}
let us compute the action. The nontrivial contributions come from the first and third terms in \eqref{Singular}. The scalar curvature is given by
\begin{eqnarray}
R=-{2\widetilde{F}^{\prime \prime}\over \widetilde{F}}-{4r(\rho)^{\prime \prime} \over r(\rho)}-{\widetilde{F}^{\prime}r(\rho)^{\prime} \over \widetilde{F}r(\rho)}-2{r(\rho)^{\prime 2}\over r(\rho)^2 } ~.
\end{eqnarray}
In the limit $\epsilon\to 0$, only the first diverges as ${\cal O}( 1/ \epsilon^2)$, which gives us the dominant contribution. One can easily check the divergence by using the following relations, 
\begin{equation}
\widetilde{F}^{\prime \prime}={\cal O}\left({ \widetilde{F}^{\prime}(\epsilon)-\widetilde{F}^{\prime}(0) \over \epsilon} \right)=-{\delta \over \epsilon} \ ,\qquad \widetilde{F}\simeq \epsilon \widetilde{F}^{\prime}(0)=\epsilon~,
\end{equation}
where we used $\widetilde{F}^{\prime}(\epsilon)=1-\delta$ and $\widetilde{F}^{\prime}(0)=1$. With these relations, consider the Einstein-Hilbert action, 
\begin{eqnarray}
S_{EH}&=&-{1\over 16\pi G} \int d \rho d \chi d\theta d \phi \sqrt{g}R=-{1\over 16\pi G} \int d \rho d \chi d\theta d \phi  \widetilde{F}(\rho)r(\rho)^2\sin \theta \left( -{2\widetilde{F}^{\prime \prime}\over \widetilde{F}}\right)\nonumber \\
&\simeq &{\pi r_h^2 \over G}\int d \rho \widetilde{F}^{\prime \prime}={\pi r_h^2 \over G} (\widetilde{F}^{\prime}(\epsilon)-\widetilde{F}^{\prime}(0))=-{4\pi r_h^2 \over 4G}\delta=-{{\cal A}_h \over 4G}\delta~,
\end{eqnarray}
where we used $r(0)\simeq r_h$ and the area of the horizon ${\cal A}_h=4\pi r_h^2$. 

The boundary contribution is given by the Gibbons-Hawking term. Doing the same way as before, we can compute the extrinsic curvature and get the dominant contribution in the limit $\epsilon \to 0$, 
\begin{equation}
K\simeq -{\widetilde{F}^{\prime}\over \widetilde{F}}+\cdots 
\end{equation}
Substituting for the Gibbons-Hawking term, we obtain
\begin{eqnarray}
S_{GH}&=&\int_{\rho=\epsilon}d^3 x \sqrt{h} {K \over 8\pi G}={1\over 8\pi G} \int d \chi d \theta d \phi \widetilde{F}r(\rho)^2 \sin \theta \left(-{\widetilde{F}^{\prime}\over \widetilde{F}} \right)  \nonumber \\
&\simeq &{-4\pi r_h^2 \over 4G}\widetilde{F}^{\prime}(\epsilon)=-{{\cal A}\over 4G}(1-\delta )~.
\end{eqnarray}
Combining the two results, we obtain the final expression for the bounce action,
\begin{equation}
S_{EH}+S_{GH}=-{{\cal A }\over 4G}~.
\end{equation}
For the case with more than one singularity, one can easily extend to  
\begin{equation}
S_{EH}+S_{GH}=-\sum_i{{\cal A }_i\over 4G}~.
\end{equation}

%
%

\end{document}